\documentstyle[prl,aps]{revtex}
\begin{document}
\draft

\twocolumn[
\hsize\textwidth\columnwidth\hsize\csname@twocolumnfalse\endcsname

\title{
   Influence of oxygen ordering kinetics on Raman and
   optical response in YBa${}_2$Cu${}_3$O${}_{6.4}$
}

\author{A.A.~Maksimov, D.A.~Pronin, S.V.~Zaitsev, and I.I.~Tartakovskii}

\address{Institute
   of Solid State Physics, Russian Academy of Sciences, Chernogolovka,
   Moscow district, 142432, Russia}

\author{G.~Blumberg,$^{1,2,4}$ M.V.~Klein,$^{1,2}$
M.~Karlow,$^{1,2}$ S.L.~Cooper,$^{1,2}$  A.P.~Paulikas,$^{3}$
and B.W.~Veal$^{3}$}

\address{
$^{1}$NSF Science and Technology Center for Superconductivity \\
$^{2}$Department of Physics, University of Illinois at
Urbana-Champaign, Urbana, Illinois, 61801-3080\\
$^{3}$Materials Science Division, Argonne National Laboratory,
Argonne, Illinois, 60439\\
$^{4}$Institute of Chemical Physics and Biophysics, R\"avala 10,
Tallinn EE0001, Estonia
}

\date{November 8, 1995; STCS-1154}
\maketitle

\begin{abstract}
Kinetics of the optical and Raman response in
YBa${}_2$Cu${}_3$O${}_{6.4}$ were studied during room temperature
annealing following
heat treatment.  The superconducting $T_c$, dc resistivity,
and low-energy optical conductivity recover slowly, implying a long
relaxation time for the carrier density. Short relaxation
times are observed for the $B_{1g}$ Raman scattering -- magnetic,
continuum, and phonon -- and the charge transfer band.
Monte Carlo simulations  suggest that these two relaxation rates are
related to two length scales corresponding to local
oxygen ordering (fast) and long chain and twin formation (slow).

\end{abstract}

]

Room temperature (RT) annealing has been observed in
oxygen deficient YBa${}_2$Cu${}_3$O${}_{6+x}$ single crystals
rapidly quenched from elevated temperatures ($T \approx 420$~K) \cite{Veal90}.
It is generally believed that the
disorder produced by heating is frozen by the fast quench, and that
RT annealing results from re-ordering
of oxygen in the CuO-chain layer via diffusion. The oxygen
re-ordering lowers the average valence
of the chain Cu and transfers holes to the CuO$_2$-planes.
Thus far, the
influence of RT annealing has been seen experimentally in
the superconducting transition temperature, $T_c$ \cite{Veal90}, the Raman
2-magnon line (2-ML) and electronic continuum intensity \cite{we94}, and
various structural \cite{Jorgensen} and optical \cite{Cardona} parameters.
However, a detailed understanding of the kinetics involved has yet
to be achieved.

We have investigated the kinetics of several physical properties of
underdoped YBa${}_2$Cu${}_3$O${}_{6.4}$ on the same sample during
reproducible quench/anneal cycles, and find two characteristic behaviors.
One is exhibited by the dc resistivity, $T_c$, and the low-frequency (LF)
optical spectral weight, all of which recover slowly over a few days
following heat treatment.
In sharp contrast, the optical
charge-transfer band (CTB), the $B_{1g}$ Raman 2-ML, electronic continuum,
and phonon intensity relax rapidly during the first hours of RT annealing.
Since the dc resistivity and $T_c$ involve the CuO$_{2}$-plane
carrier density directly, our results
suggest that the rapid relaxation of the second group occurs through a process
that is independent of chain-to-plane charge transfer.

Monte Carlo simulations of oxygen
ordering in the chains also show two relaxation rates: one fast,
corresponding to short range oxygen order where the correlation length
is of order the lattice constant, $r \approx a$; and one slow, where
the correlation length is large compared to $a$.
We conclude from these results that the fast relaxation processes depend on
short
range correlations, whereas long range oxygen order is important to the
slower processes.

As shown in the inset of Fig.~\ref{T_c}, the underdoped
YBa${}_2$Cu${}_3$O${}_{6.4}$ single crystal used for these
measurements displays a sharp superconducting transition ($\Delta
T_c \lesssim 1 \mbox{--} 1.5$~K) at each quench/anneal cycle, despite
$T_c$ changes from 18~K before the thermal treatment to
2.9~K immediately following the quench. The reproducibility of these results
for many cycles indicates that the total oxygen content of the
sample remains constant.  The homogeneity of the oxygen distribution
at the 1~$\mu$m level was confirmed by micro Raman experiments.

The optical reflectance at near-normal incidence in the energy range
 $\omega = $ 0.03--2.25~eV was measured at successive times
 following thermal treatment using a rapid-scan Michelson interferometer.
A Kramers--Kronig transformation \cite{stern} was applied
 to obtain the optical conductivity $\sigma(t,\omega)$ from the
 measured reflectance data.
Since variations in the high frequency reflectance ($3.8 <
\omega < 4.3$~eV)\cite{Cardona} cause changes of less then 5\% in the
conductivity $\sigma(t,\omega)$ for $\omega < 2.5$~eV, ellipsometric data
measured after complete relaxation were used in the energy range $\omega =
2.25 \mbox{--} 6$~eV to make this transformation.
The inset of Fig.~\ref{optics} shows
 $\sigma(t,\omega)$ for YBa${}_2$Cu${}_3$O${}_{6.4}$ at two times $t$
 following the quench, $t_0 = 10$~min and $t_{\infty}$ (fully relaxed).
Immediately after the quench, the largest changes in
$\sigma(t,\omega)$
 are a decrease in the LF spectral weight (box 1 in the inset),
 and an increase in the CTB spectral weight (box 2) \cite{cooper}.
 The LF spectral weight (box 1) is comprised of a Drude part
 due to holes in the CuO${}_2$-planes plus a
 mid-IR peak associated with chain conductivity \cite{Widder}.
In Fig.~\ref{optics} we plot as a function of time the variation in
 effective number of carriers (integrated spectral weight), $\Delta
N_{1,2}(t) = C \int_{box\, 1,2}[\sigma(t,\omega) -
\sigma(t_{\infty},\omega)] \, d\omega$, where the constant $C$ is given in
Eq.~(3) of Ref.~\onlinecite{cooper}.  We find that
$\Delta N_{2}(t)$ takes $\approx 100$~min to recover to its initial value,
whereas complete recovery of $\Delta N_{1}(t)$ requires several days.
Moreover, the sum $\Delta N(t) = \Delta N_{1}(t) + \Delta N_{2}(t)$ during
 the first $\sim$~100~min remains approximately constant.
These results
 are significant because they suggest that the transfer of spectral weight
 from the CTB to LF band occurs quite rapidly relative to the time scale
 associated with chain-to-plane charge transfer.  Notably, the monotonic
increase in $\Delta N_{1}(t)$ at long times may reflect different effective
masses
for holes in the chain and the plane layers.

Raman scattering (RS) data from single crystal
YBa${}_2$Cu${}_3$O${}_{6.4}$
was obtained in 20 second intervals at successive times following
the quench using a CCD equipped Raman triple spectrometer.
The 2-ML \cite{l87,b94} and electronic continuum
\cite{b94,rez93} spectra were obtained at room
temperature, while the 336~cm$^{-1}$ phonon scattering data were
taken at 5~K.
The inset of Fig.~\ref{Raman} presents the RS spectrum measured at $t_{0}=
6$~min after the quench in the $x'y'$ (mainly $B_{1g}$) geometry.
Fig.~\ref{Raman} shows the temporal dependence of the 2-ML and electronic
background (0.1 to 0.5~eV) intensities.
An increase of both the 2-ML and
 low-frequency RS background intensities is
 observed after quenching; however the fractional increase
 is larger for the background than for the 2-ML.
Interestingly, the main change in the intensity for both features
 occurs during the initial 50--100~min, although complete recovery
 to pre-quench values takes several days.

At least two characteristic relaxation
 rates may be extracted from Figs.~1--3 if an exponential
 temporal dependence of the
 form $A(t)-A(t_\infty)=[A(t_0)-A(t_\infty)]\exp(-t/\tau)$ is assumed.
 In order to estimate these times, we have represented the data as
 \begin{equation}
 L_A(t) \equiv
 \ln{\left|\frac{A(t)-A(t_{\infty})}{A(t_0)-A(t_{\infty})} \right|},
 \end{equation}
 where $A(t)$ is one of the measurable quantities and $A(t_0)$ is the
 first measured value of $A$ just after the quenching.

Figure ~\ref{L_A}a plots $L_A(t)$ for the optical and RS quantities,
and $T_c$.
Two distinct relaxation behaviors are clearly seen.
The kinetics of $T_c$ and $\Delta N_{1}(t)$ demonstrate
a single long relaxation time, estimated to be $\tau \approx$~250~min
via a straight line least squares fit to the experimental points in the
range $t=0 - 200$~min.
In \cite{VealResist}, the kinetics of the dc
resistivity and $T_c$ during room temperature
 annealing were shown to be of the same order.
By contrast, the kinetics of 2-ML, background intensity, the
 integrated intensity of the $B_{1g}$ phonon
 and $\sigma(t,\omega)$ in the
 CTB spectral region demonstrate two relaxation times: an initially short
 relaxation ($\tau \approx 50$~min) over the first $ \approx 100$~min,
 followed by a considerably slower relaxation.

To gain insight into the experimental data, we carried out a
 Monte Carlo simulation of the oxygen distribution kinetics in the
 CuO-chains of YBa${}_2$Cu${}_{3}$O${}_{6+x}$.
We used the asymmetric next-nearest-neighbor Ising model
 \cite{Fontaine1990}, which takes into account three different effective
 pair interactions between oxygen atoms  ($V_1 = 367$~meV, $V_2 = -0.348 V_1$
 and $V_3 = 0.159 V_1$) \cite{PAA1991a}.
This model has been used previously \cite{PAA1991a,Fontaine1991} for
 collecting snapshots of oxygen placement in the CuO-chains at each Monte Carlo
 step to study the kinetics of oxygen ordering, and also as a method
 for determining final oxygen structure \cite{Fontaine1992,PAA1991b}.
We parameterize the results of these simulations in terms of an ordering
parameter $\alpha$, which measures how close the structure is to the ordered
Ortho-I (OI)-phase \cite{MC-later}.
$\alpha$ is defined to be
zero when the structure is fully disordered and unity when the structure
is in the ideal OI-phase.
We calculate $\alpha$ for different correlation distances $r$ by
comparing, at every oxygen position, the Monte Carlo oxygen arrangement
on a square of side $2r$ to the perfect OI structure.
We used up to 5000 Monte Carlo steps on a
 128$\times$128 lattice at room temperature, starting with a
 disordered structure.
To compare the calculated and experimental data we defined function
 $L_\alpha(t)$, for the ordering parameter $\alpha$ that is similar
 to $L_A(t)$ (Eq.~1), where $t_\infty$ is equal to 5000 Monte Carlo steps.
Fig.~\ref{L_A}b shows the time dependence of $L_\alpha$
 for different correlation distances $r$ with oxygen content $x = 0.4$.
It is seen that the dependence of $L_\alpha$ on time
 for different $r$ looks like the temporal dependence of the various
 measured quantities.
For long correlation distances $L_\alpha$
 demonstrates slow kinetics, whereas the time dependencies of $L_\alpha$ for
 short distances ($r=$ 1 -- 2) demonstrate two rates: fast for the initial
 stage and much slower at longer times.
The appearance of two relaxation rates at small $r$ results from the
 fact that at the initial stage of relaxation only local ordering
 takes place.
Consequently, many little twins (a few lattice constants in size) appear.
This process is rather fast and the associated correlation lengths $r$
are quite small.
The slower kinetics correspond to motion of twin walls which results in the
subduction of smaller twins by larger ones.
This process is revealed at all correlation distances \cite{MC-later}.
Similar results hold for the Ortho-II ordering parameter.
As expected, calculations performed for higher values of the oxygen
content show less difference between the kinetics at various
correlation distances.

The long relaxation times exhibited by the dc conductivity \cite{VealResist},
 $T_c$ and $\Delta N_{1}(t)$ imply that
 the fast kinetic behavior displayed by the CTB and the $B_{1g}$
 RS features
 occurs before oxygen ordering in the chains causes significant hole transfer
 to the CuO$_2$-planes.
Since some rapidly annealing parameters, such as the 2-ML,
 are believed to have their origins in the planes, {\em we conclude
 that short range oxygen-ordering in the chains gives rise to
 short-length-scale changes in the planes without significantly
 altering the average planar carrier density.}

Comparing Fig.~\ref{L_A}a with Fig.~\ref{L_A}b, one can conclude that
the kinetics for $\Delta N_{2}(t)$ and for the $B_{1g}$
 symmetry RS are determined by local oxygen ordering
 at nearest neighbor distances.
As discussed in \cite{cooper}, the CTB is most probably due to a
local excitation of a hole in the
 CuO$_2$-planes from Cu(2) $d_{x^2-y^2}$ to the four O(2,3) $p_{x,y}$
 states that surround a first-neighbor Cu(2) site.
The rapid transfer of the integrated spectral weight from $\Delta
N_{2}(t)$ to $\Delta N_{1}(t)$ suggests partial delocalization of bound
carriers.

 The 2-ML mainly probes short wavelength magnetic excitations
 from the vicinity of the magnetic Brillouin zone boundary
 ${\bf k} = (\pi/a,0)$, which occur via photon-driven superexchange
 between two nearest-neighboring Cu(2) spins in the antiferromagnetic
 surroundings.
Thus, the CTB and 2-ML both require only short range anti-ferromagnetic
 order which is affected by changes in the local carrier distribution
 but is independent of the average carrier number.
The rather fast kinetics of the Raman $B_{1g}$ continuum implies that it
is also a local excitation, perhaps of magnetic origin,
 {\em and that it is not related directly to the average
 carrier number.}
Phonon interaction with the continuum
might explain the observed phonon line intensity kinetics through a
Fano-type effect, but a detailed model needs to be formulated to
understand this.
By contrast, the resistivity, $T_c$, and the LF absorption kinetics
are determined by longer distance oxygen ordering connected with the
 formation of ordered oxygen chains several lattice constants long.
The LF band has been shown to be polarized along the chain direction
 \cite{cooper}; thus chain fragments of a minimum length are
 required for this band to occur.

In summary, we have measured the kinetic behavior of the
magnetic susceptibility, the $B_{1g}$ symmetry Raman scattering
 (2-ML, electronic continuum and the 336~cm$^{-1}$ phonon), and the
 optical conductivity (LF and CTB) of YBa${}_2$Cu${}_3$O${}_{6.4}$
 single crystals during room temperature annealing.  The kinetics of the Raman
features and the optical conductivity in the CTB region demonstrate fast
 relaxation ($\tau_f \sim 50$~min).  The relaxation time for the
resistivity, the  LF absorption, and $T_c$ is much larger ($\tau_s \gtrsim
 220$~min), implying that there is a nearly constant average carrier number
 for times less than $\tau_f$.  Comparing the measured kinetics with Monte
Carlo model simulations we suggest that the short-length-scale excitations
 are responsible for the fast processes, such as $B_{1g}$ symmetry RS and
 the CTB in the optical conductivity.  Longer correlations ($r > 2$) are
responsible for changing the resistivity, LF optical conductivity and hence
the effective carrier concentration.

The joint Russian/U.S. work was supported by NATO CRG No. 92-1239, the work at
Chernogolovka by RBRF 95-02-06111a, ISF NKW000 and
RSP on HTSC 93193, the work at Urbana
by NSF DMR 91-20000 through STCS, and the
work at Argonne by US DOE  W-31-109-ENG-38.

\begin{figure}
\caption{
Temporal dependencies of $T_c$ after the quench.
{\em  Inset}: Dependence of
magnetic susceptibility $\chi$ on temperature measured at different
times after the quench.
}
\label{T_c}
\end{figure}

\begin{figure}
\caption{
Temporal dependencies after the quench of the variation in
the optical
conductivity integrated over the LF (open circles) and CTB (filled
circles) regions.
{\em Inset}: Optical conductivity $\sigma(t,\omega)$
before the heat treatment and 10~min after the quench. Boxes 1 and 2 show
the integration regions.
}
\label{optics}
\end{figure}

\begin{figure}
\caption{
Temporal dependencies of the intensity of the
2-magnon Raman scattering line maximum (open circles) and the background
intensity (filled circles) after the quench.
The  vertical axis scales are anchored to the time axis at their
pre-quench values (i.e., 7.7 for background and 38.5 for 2-ML).
{\em Inset}: Raman spectrum measured in $x^{\prime}y^{\prime}$ geometry
at $t= 6$~min after the quench. Filled region corresponds to broad
structureless electronic background.
}
\label{Raman}
\end{figure}

\begin{figure}
\caption{
(a) - Measured values $L_A(t)$ [Eq.~(1)] for parameters in Figs.~1--3.
Straight lines are the results of least squares fitting of
experimental data at the initial stage of aging.
Crosses - $T_c$, $\tau \approx 340$~min;
squares - $\Delta N_{1}(t)$, $\tau \approx 220$~min;
filled circles - $\Delta N_{2}(t)$, $\tau \approx 59$~min;
triangles - background, $\tau \approx 51$~min;
open circles - 2-magnon line, $\tau \approx 36$~min.
\protect\\
\hspace*{15.2mm}(b) - Time dependencies of $L_{\alpha}$ for
calculated parameter $\alpha$ in YBa${}_2$Cu${}_3$O${}_{6.4}$ crystals for
different
correlation distances $1 \leq r \leq 6$ in units of $\protect\sqrt{2}a$.}
\label{L_A}
\end{figure}

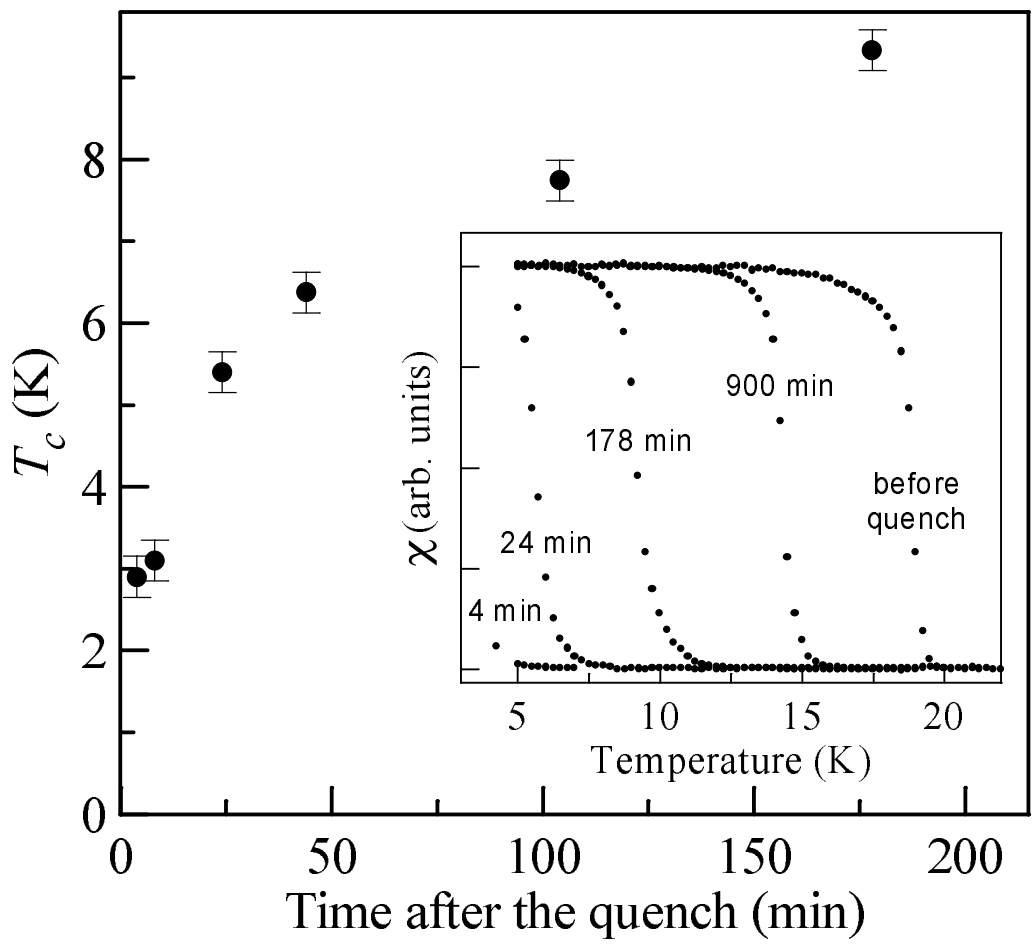
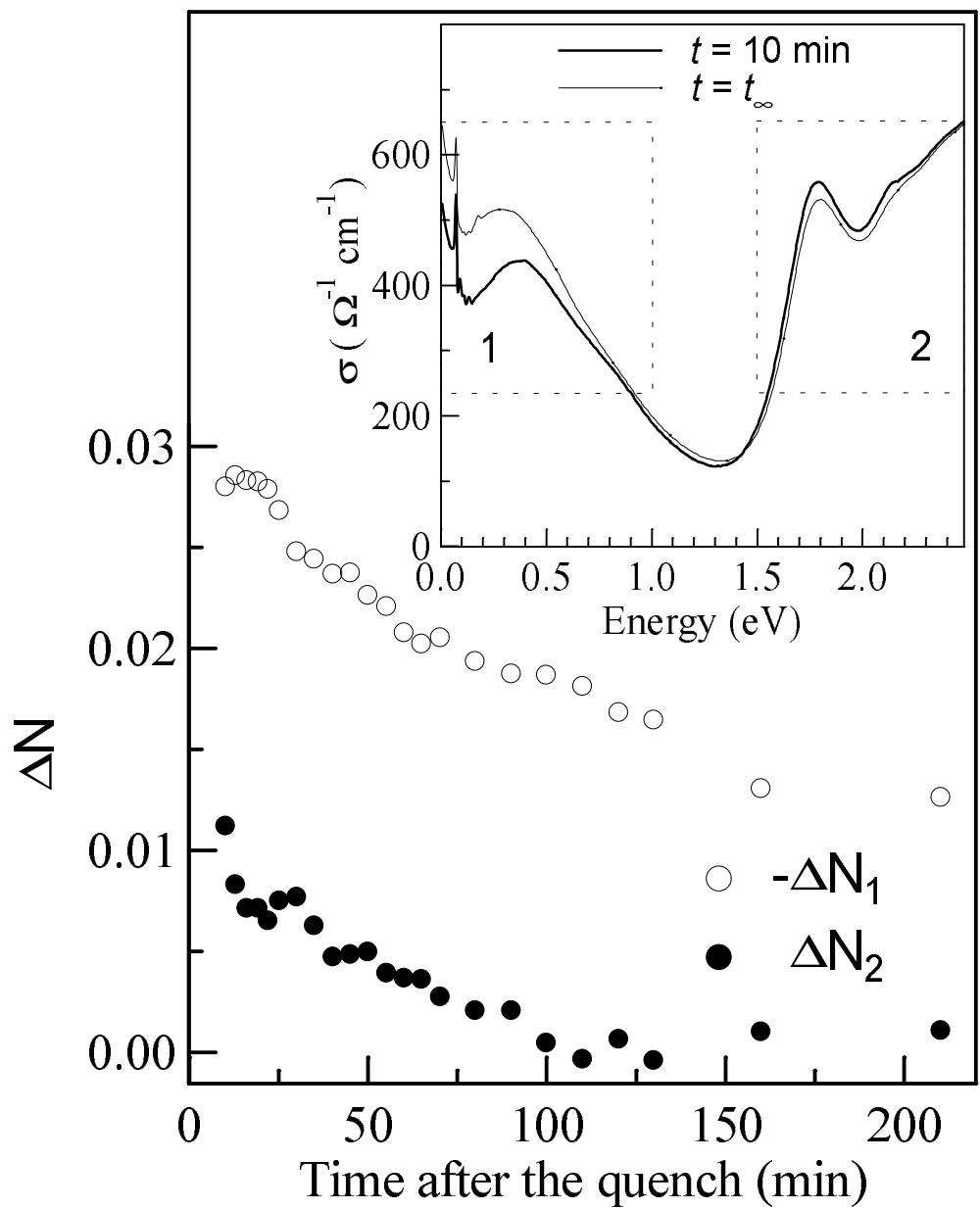
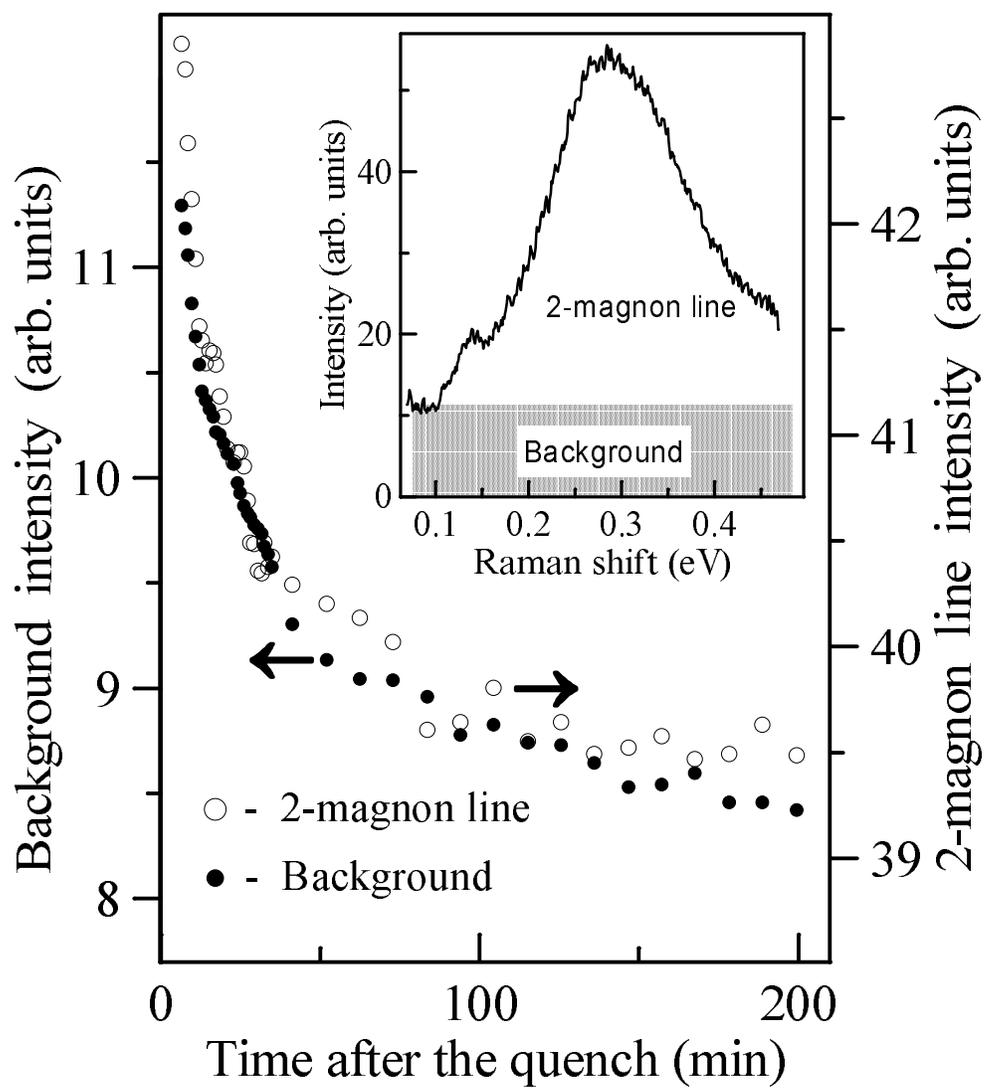
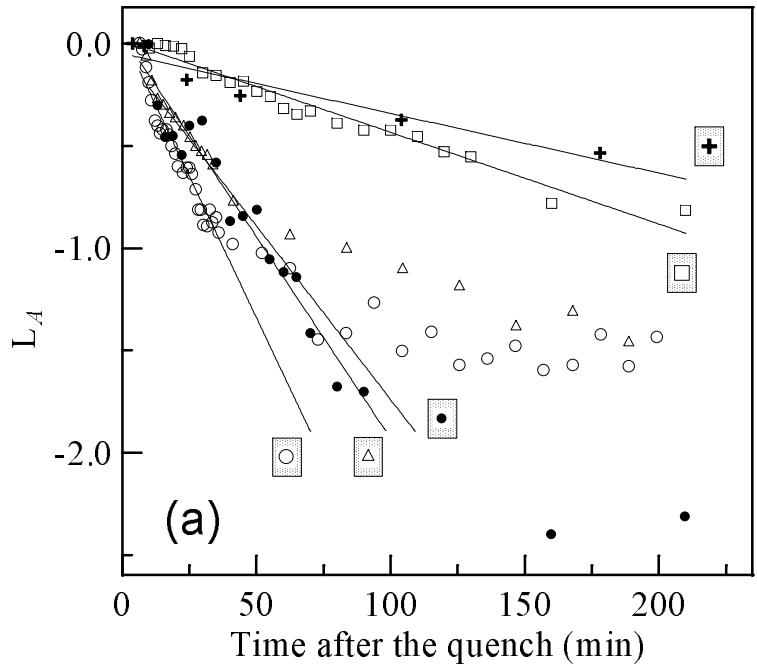
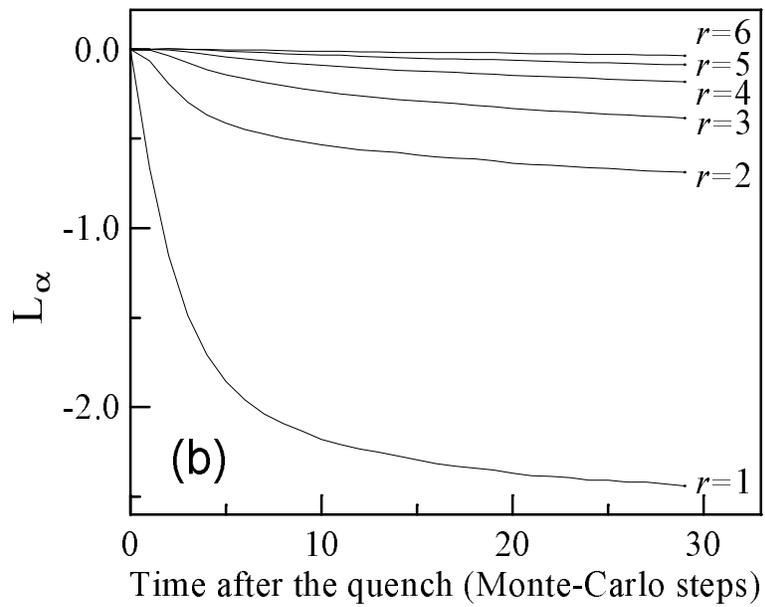
\end{document}